\documentclass[referee]{aa}
\usepackage{graphicx}

\begin{document}
\title{Galaxy Flow in the Canes Venatici\,I Cloud
\thanks{Based on observations made with the NASA/ESA Hubble Space
Telescope.  The Space Telescope Science Institute is operated by the
Association of Universities for Research in Astronomy, Inc. under NASA
contract NAS 5--26555.}}
\titlerunning{Canes Venatici cloud }
\author{I. D. Karachentsev \inst{1}
\and M. E. Sharina \inst{1,10}
\and A. E. Dolphin \inst{2}
\and E. K. Grebel \inst{3}
\and D. Geisler \inst{4}
\and P. Guhathakurta \inst{5}
\and P.~W. Hodge \inst{6}
\and V. E. Karachentseva \inst{7}
\and A. Sarajedini \inst{8}
\and P. Seitzer \inst{9}}
\institute{Special Astrophysical Observatory, Russian Academy
of Sciences, N. Arkhyz, KChR, 369167, Russia
\and Kitt Peak National Observatory, National Optical Astronomy
Observatories,
P.O. Box 26732, Tucson, AZ 85726, USA
\and Max-Planck-Institut f\"{u}r Astronomie, K\"{o}nigstuhl 17, D-69117
Heidelberg, Germany
\and Departamento de F\'{\i}sica, Grupo de Astronom\'{\i}a, Universidad de
Concepci\'on, Casilla 160-C, Concepci\'on, Chile
\and UCO/Lick Observatory, University of California at Santa Cruz, Santa
Cruz, CA 95064, USA
\and Department of Astronomy, University of Washington, Box 351580,
Seattle, WA 98195, USA
\and Astronomical Observatory of Kiev University, 04053, Observatorna 3,
Kiev, Ukraine
\and Department of Astronomy, University of Florida, Gainesville, FL
32611, USA
\and Department of Astronomy, University of Michigan, 830 Dennison
Building, Ann Arbor, MI 48109, USA
\and Isaac Newton Institute, Chile, SAO Branch}
\date{Received:  August 8, 2002}
\abstract{
We present an analysis of
Hubble Space Telescope/WFPC2 images of eighteen galaxies in
the Canes Venatici I cloud. We derive their distances from the luminosity
of the tip of the red giant branch stars with a typical accuracy of
$ \sim 12$ \%.  The resulting distances are
3.9 Mpc (UGC 6541), 4.9 Mpc (NGC 3738), 3.0 Mpc (NGC 3741),
4.5 Mpc (KK 109), $>6.3$ Mpc (NGC 4150), 4.2 Mpc (UGC 7298), 4.5 Mpc
(NGC 4244), 4.6 Mpc (NGC 4395), 4.9 Mpc (UGC 7559), 4.2 Mpc (NGC 4449),
4.4 Mpc (UGC 7605), 4.6 Mpc (IC 3687), 4.7 Mpc (KK 166),
4.7 Mpc (NGC 4736), 4.2 Mpc (UGC 8308), 4.3 Mpc (UGC 8320),
4.6 Mpc (NGC 5204), and 3.2 Mpc (UGC 8833). The CVn\,I cloud has a mean
radial velocity of 286 $ \pm$ 9 km~s$^{-1}$, a mean distance of 4.1 $ \pm$ 0.2 Mpc,
a radial velocity dispersion of 50 km~s$^{-1}$, a mean projected radius of
760 kpc, and a total blue luminosity of $2.2 \cdot 10^{10} L_{\sun}$. Assuming virial
or closed orbital motions for the galaxies, we estimated their virial and
their orbital
mass-to-luminosity ratio to be 176 and 88 $ M_{\sun}/L_{\sun} $, respectively.
However, the CVn\,I cloud is characterized by a crossing time of 15 Gyr,
and is thus far from a state of dynamical equilibrium. The large crossing
time for the cloud, its low content of dSph galaxies ($ < 6$ \%),
and the almost ``primordial'' shape of its luminosity function show that the
CVn\,I complex is in a transient dynamical state, driven rather by the
free Hubble expansion than by galaxy interactions.
}
\maketitle

\section{Introduction}

  About 1/7 of the 240 known galaxies with radial velocities
$V_{LG} < 400$ km~s$^{-1}$ are concentrated in a small area in the Canes Venatici
(CVn) constellation  [$\alpha = 11^h 30^m$ to $13^h 40^m$, $\delta = +25\degr$ to $+55\degr$]
which occupies only 1/50 of the sky. Therefore, the apparent
overdensity of the number of galaxies seen in the CVn direction exceeds
$\delta N/ \langle N\rangle \sim 7$. This scattered complex of nearby galaxies has been
noted by many  authors (Karachentsev 1966; de Vaucouleurs 1975; Vennik
1988). In the Nearby Galaxies Catalog (Tully 1988) the group is indicated
by ``14---7'' (CVn\,I) as a part of the more extended Coma-Sculptor
cloud, containing the Local Group (LG) and also the M81, Cen~A, and Sculptor
galaxy groups. In contrast to the groups mentioned, the CVn\,I cloud is
populated mostly by late-type galaxies of low luminosity.

  At present the structure and kinematics of the CVn\,I complex
are still poorly understood because of the lack of reliable data on the galaxy
distances. Sandage \& Tammann (1982) determined the distance to IC 4182
(4.70 Mpc) from the luminosity of Cepheids. Distance estimates
from the luminosity of the brightest stars were derived for DDO 154
(Carignan \& Beaulieu 1989), DDO 168 (Bresolin et al. 1993) and UGC 8508
(Karachentsev et al. 1994). Using this method, Georgiev et al. (1997),
Makarova et al. (1997, 1998), Karachentsev \& Drozdovsky (1998), and
Sharina et al. (1999) determined distance moduli for 35 spiral and
irregular galaxies in CVn\,I. The median distance to the cloud was found
to be 4.3 Mpc, which is in good agreement with the single Cepheid distance
estimate. However, considerable distance modulus errors ($ \sim$ 0.5 mag)
have hampered the study of structure and kinematics of the CVn\,I complex. That
is why the CVn\,I objects were included in the program of our snapshot
survey of nearby galaxies with the Hubble Space Telescope (Seitzer et al.
1999; Grebel et al.\ 2000), where galaxy distances are determined on the basis of a much more precise
method, via the luminosity of red giant branch tip stars.  In the
framework of this study of galaxies in the Local Volume we earlier 
investigated the Centaurus A group (Karachentsev et al.\ 2002a) and
the M81 group (Karachentsev et al.\ 2000, 2001, 2002b).  The first
distance measurements for five members of the Cloud based on the HST data
have already been published
(Karachentsev et al. 2002c). Here, we present new distances for 18
galaxies in the CVn\,I area.

  \section{ WFPC2 photometry}

   The galaxy images were obtained with the Wide Field and Planetary
Camera (WFPC2) aboard the Hubble Space Telescope (HST) between July 23,
1999 and June 20, 2001 as part of our HST snapshot survey (proposals GO 8192,
8601) of nearby galaxy candidates (Seitzer et al. 1999). Each galaxy was 
observed in the F606W and F814W filters (one 600~s exposure in each filter).
Digital Sky Survey images of the galaxies are shown in Figure 1 with the HST 
WFPC2 footprints superimposed. Small galaxies were usually
centered on the WF3 chip.  For some bright objects the WFPC2
position was shifted towards the galaxy periphery to decrease 
stellar crowding. The WFPC2 images of the galaxies are
presented in the upper panels of Figure 2, where both filters are combined.
The compass in each field indicates the North and East directions.

 For photometric measurements we used the HSTphot stellar photometry
package developed by Dolphin (2000a). The package has been optimized
for the undersampled conditions present in the WFPC2 to work in
crowded fields.  After removing cosmic rays, simultaneous
photometry was performed on the F606W and F814W frames using
\textit{multiphot} procedure. Resulting instrumental magnitudes of
radius $0\farcs5$ were corrected for charge-transfer inefficiency and
converted to standart V and I magnitudes  using the relations (11) and (12) of Dolphin (2000b).
This calibration equations are analogous to equations of Holtzman et al. (1995),
but incorporate the pixel area corrections.
 Additionally, stars with a signal-to-noise ratio $S/N < 3 $,
$\mid \chi \mid\,\, >2.0$, or  $\mid$ sharpness $\mid\,\, >0.4$  in either
exposure were eliminated from the final photometry list. The uncertainty
of the photometric zero point is estimated to be within $0\fm05$
(Dolphin 2000b).

\section{ CMDs and galaxy distances}

During the last decade the tip of the red giant branch (TRGB) method has
become an efficient tool for measuring galaxy distances.
The TRGB distances agree with the distances derived from the Cepheid
period-luminosity relation within a 5\% error. As was shown by
Lee et al. (1993), Salaris \& Cassisi (1997), and Udalski et al. (2001),
in the $I$ band the TRGB position is relatively independent of age
and metallicity within $ \sim 0.1$ mag  for old stellar population
with [Fe/H] $< - 0.7$ dex.  According to
Da Costa \& Armandroff (1990), for metal-poor systems the TRGB
is located at $M_I = -4.05$ mag. Ferrarese et al. (2000) calibrated the
zero point of the TRGB from galaxies with Cepheid distances and yielded
$M_I = -4.06 \pm 0.07$. A new TRGB calibration $M_I = -4.04 \pm0.12$ mag
was determined by Bellazzini et al. (2001) based on photometry and
distance estimation from a detached eclipsing binary in the Galactic
globular cluster $ \omega$ Centauri. In the present paper we adopt
$M_I = -4.05$ mag. The bottom left panels of Figure 2 show
$I$ versus $(V-I)$ color-magnitude diagrams (CMDs) for the eighteen
observed galaxies.

   We determined the TRGB location using a Gaussian-smoothed $I$-band
luminosity function (LF) for red stars with colors $(V-I)$ within $\pm0\fm5$
of the mean $<V-I>$ for expected red giant branch stars. Following
Sakai et al. (1996), we applied a Sobel edge-detection filter.
The position of the TRGB was identified with the peak in the
filter response function. The resulting LFs and the Sobel-filtered LFs
are shown in the bottom right corners of Figure 2. The results are
summarized in Table 1. This table contains the following columns:
(1) galaxy name; (2) equatorial coordinates corresponding to the
galaxy center; (3,4) angular size in arcmin and apparent total magnitude
from the NASA Extragalactic Database (NED), uncorrected for internal and
external extinction;(5) Galactic extinction in the $B,I$- bands from
Schlegel et al. 1998; (6) integrated or effective color of the galaxy,
corrected for Galactic extinction; the data on colors are taken from
Makarova et al. (1998), Makarova (1999), and Prugniel \& Heraudeau (1998);
for the galaxies NGC~3741, KK~109, and KK~166 we present $V-I$ colors of their
core from our measurements; (7) morphological type
in de Vaucouleurs notation; (7) radial velocity with respect to the
LG centroid; for some galaxies we used new accurate velocities measured
recently by Huchtmeier et al. (2002); (9) position of the TRGB and its
uncertainty derived with the Sobel filter; (10) true distance modulus
with its uncertainty, which takes into account the uncertainty in the TRGB
detection as well as uncertainties of the HST photometry zero point
($\sim0\fm05$), the aperture corrections ($\sim0\fm05$), and crowding effects
($\sim0\fm06$) added in quadrature; uncertainties in extinction and
reddening are assumed to be $10\%$ of their values from Schlegel et al.(1998);
and (11) linear distance in Mpc and its uncertainty.

   Given the distance moduli of the galaxies, we can estimate their
mean metallicity, [Fe/H], from the mean color of the TRGB measured at an
absolute magnitude $ M_{I} = -3.5$, as recommended by Da Costa \& Armandroff
(1990). Based on a Gaussian fit to the color distribution of the giant
stars in a corresponding $I$- magnitude interval $ (-3.5\pm 0.3)$, we derived
their mean colors, $ (V - I)_{-3.5} $, which lie in a range of [1.15 - 1.68]
after correction for Galactic reddening. Following the relation of Lee et al.
(1993), this provides us with mean metallicities, [Fe/H] = [ -0.8, -2.5] dex,
listed in the last column of Table 1. With a typical statistical scatter
of the mean color ($\sim0\fm05$), and uncertainties of the HST photometry
zero point we expect an uncertainty in metallicity to be about 0.3 dex.
Therefore within the measurement accuracy the metallicity of the galaxies
satisfy the required limitation, [Fe/H] $< -0.7$ dex.

Below, some individual properties of the galaxies are briefly discussed.

{\em UGC 6541 = Mkn 178.} A blue compact galaxy from Markarian's lists is
located on the NW edge of the CVn\,I cloud. It was resolved into stars
for the first time by Georgiev et al.(1997), who derived a distance of
3.5$\pm 0.7$ Mpc via the brightest blue stars. The distance to UGC 6541
from the luminosity of TRGB is D = 3.89$\pm0.47$ Mpc, which is in reasonable
agreement with the previous estimate.

{\em NGC 3738 = UGC 6565 = Arp 234.} This dwarf irregular galaxy appears
to be semi- resolved into brightest stars on the reproduction given in the
Atlas of Peculiar Galaxies (Arp 1966). Georgiev et al.(1997) estimated
its distance as 3.5$\pm 0.7$ Mpc from the magnitudes of the brightest blue
stars. The images obtained with WFPC2 reveal about 17700 stars seen in
both filters. The CM diagram for NGC 3738 shows a large number of blue stars,
as well as AGB stars. From the TRGB position we derive a 
distance of 4.90$\pm 0.54$ Mpc.

{\em NGC 3741 = UGC 6572.} Like two previous objects, NGC 3741 lies
at the NW periphery of the CVn\,I cloud. The galaxy has an asymmetric
``cometary'' shape. Its size in H\,{\sc i} exceeds its optical diameter
significantly (Haynes \& Giovanelli 1991). Georgiev at al. (1997)
derived its distance to be 3.5$\pm 0.7$ Mpc from the brightest blue
stars, while the TRGB distance from
our measurements is 3.03$\pm 0.33$ Mpc.

{\em KK 109.} This dwarf irregular galaxy of low surface brightness was
found by Karachentseva \& Karachentsev (1998). Huchtmeier et al. (2000)
detected it in the H\,{\sc i} line and determined its radial velocity, $V_{LG} =
241$ km~s$^{-1}$, which is typical for CVn\,I members. 
The CM diagram of KK 109 shows
the TRGB magnitude to be 24.26$\pm 0.15$ mag, which yields a distance of
4.51$\pm 0.34$ Mpc. Based on the WFPC2 images, we carried out surface
photometry of KK 109, obtaining total magnitudes $V_t = 18.14 \pm 0.2$
mag, $(V - I)_t = 0.8 \pm 0.1$ mag, and a central surface brightness of
$23.6 \pm 0.2$ mag arcsec$^{-2}$ in the $V$ band. With the derived apparent
magnitude and distance, KK 109 has an absolute magnitude of $M_V =
-10.19$ mag, placing it among the faintest known dIrr galaxies such as 
LGS\,3 and Antlia.

{\em NGC 4150.} The core of this lenticular galaxy is crossed by a curved
dusty furrow (see insert in Fig. 2). In spite of its low radial velocity,
$V_{LG} = 198$ km~s$^{-1}$, NGC 4150 appears unresolved into stars on the WFPC2 images.
Its TRGB magnitude appears to exceed $I_{lim} = 25$ mag (beyond our
detection limit), yielding a lower limit of 6.3 Mpc for its distance. 
We suggest that NGC 4150 belongs to the Virgo cluster outskirts, and not
to the CVn\,I cloud. Most of the objects seen in the galaxy body seem to be
slightly extended and diffuse with integrated colors of $V-I$ = 0.8 - 1.6,
which raises the possibility that they are globular
clusters.  If they are indeed globular clusters, we can use the turnover
magnitude of the globular cluster luminosity function (GCLF), $V\sim 24$ mag,
as a distance indicator (Ferrarese et al. 2000).  With this assumption we derive
a rough distance estimate of $\sim 20 $ Mpc consistent with the Virgo cluster
distance.

{\em UGC 7298.} This is a dIrr galaxy of low surface brightness.
UGC 7298 has been resolved into stars by Tikhonov \& Karachentsev (1998),
who estimated its distance to be 8.6$\pm 1.5$ Mpc via the brightest stars.
The CM diagram (Fig. 2) shows populations of blue stars and AGB stars.
We determine the TRGB magnitude to be $I$(TRGB) = 24.12 $\pm 0.15$ mag, which
gives a distance of 4.21$\pm 0.32$ Mpc. The TRGB distance is approximately
two times smaller than the distance from the brightest stars.  A possible
cause of this difference is a lack of very luminous blue stars in this
galaxy, i.e., no very recent massive star formation.

{\em NGC 4244.} A large edge-on Sc galaxy extends far beyond the WFPC2
field. Its periphery was resolved into stars by Karachentsev \& Drozdovsky
(1998), who estimated the galaxy distance to be 4.5 $\pm 0.9$ Mpc from the photometry
of the brightest stars. The CM diagram (Fig. 2) shows $\sim15000$ stars, 
in particular pronounced 
populations of blue stars and AGB stars. The TRGB position, 24.25$\pm 0.22$
mag, corresponds to a distance of 4.49$\pm 0.47$ Mpc in close agreement
with the distance estimate via the brightest stars.

{\em NGC 4395.} This face-on Sd galaxy with a Seyfert 1 type nucleus also
extends beyond the WFPC2 field. According to Karachentsev \& Drozdovsky
(1998), its distance via the brightest blue stars is 4.2$\pm 0.8$ Mpc. The
CM diagram in Fig. 2 reveals about 21800 stars seen both in the $V$ and $I$ bands.
The majority of the detected stars are likely RGB stars.
>From the TRGB position we derive a distance of 4.61$\pm 0.57$ Mpc, which
agrees well with the previous distance estimate.

{\em UGC 7559 = DDO 126}. This irregular dwarf galaxy has been resolved
into stars by Hopp \& Schulte-Ladbeck (1995), Georgiev et al. (1997), and
Makarova et al. (1998), who derived distance estimates of 4.8 Mpc, 3.9 Mpc,
and 5.1 Mpc, respectively. Our distance for UGC 7559 based on the TRGB
(4.87$\pm 0.55$ Mpc) is in the middle of the range previously obtained by
other authors.

{\em NGC 4449.} This boxy-shaped Magellanic irregular galaxy of high
surface brightness is a second ranked member of the CVn\,I cloud according
to its luminosity. NGC 4449 is enveloped in a huge H\,{\sc i} ``fur coat,'' whose
angular size ($75'$) exceeds the Moon's diameter (Bajaja et al. 1994).
Based on photometry of the brightest stars, Karachentsev \& Drozdovsky
(1998) estimated its distance to be 2.9$\pm 0.6$ Mpc. The WFPC2 photometry
reveals about 27000 stars seen in both images. The CM diagram in
Fig. 2 shows stellar populations of different kinds including 
RGB stars. From the TRGB magnitude, 24.11$\pm 0.26$ mag, we derive a distance
of 4.21$\pm 0.50$ Mpc.

{\em UGC 7605.} This is a blue irregular galaxy shaped like a horseshoe.
The brightest blue stars in UGC 7605 are concentrated towards the core,
and the outlying galaxy parts are redder and smooth. From the luminosity
of the brightest stars Makarova et al. (1998) derived a galaxy distance
of 4.4$\pm 0.9$ Mpc. The CM diagram of UGC 7605 (Fig. 2) shows the RGB
population giving a distance of 4.43$\pm 0.53$ Mpc, which is in close agreement
with the previous estimate.

{\em IC 3687 = DDO 141 = UGC 7866.} IC 3687 is an irregular dwarf galaxy
with several regions of current star formation activity. Its CM
diagram shows a mixed stellar population with a pronounced RGB 
with $I$(TRGB) = 24.29$\pm 0.22$ mag, which corresponds to a distance of
4.57$\pm 0.48$ Mpc. Our distance for IC 3687 differs from the previous
distance, 3.0$\pm 0.6$ Mpc, obtained by Makarova et al. (1998) via the
brightest stars.

{\em KK 166.} This galaxy is unique in terms of being the only
dwarf spheroidal (dSph) galaxy of very low surface
brightness identified so far in the CVn\,I region. (Another possible 
dSph in CVn\,I is DDO 113 = KDG 90). The galaxy has been observed but
not detected in the H\,{\sc i} line by Huchtmeier et al. (2000). The CM diagram
shows a dominant RGB population with $I$(TRGB) = 24.36$\pm 0.34$ mag,
yielding a distance of 4.74$\pm 0.69$ Mpc, which confirms KK 166 as a
likely member of the CVn\,I cloud. Apart from stellar photometry, we also carried
out surface photometry in circular apertures. From our
measurements KK 166 has a total magnitude  $V_t = 16.8 \pm 0.2$ mag,
$(V-I)_t = 1.2 \pm 0.1$ mag, and a central surface brightness of 25.0$\pm
0.2 ^m/\sq\arcsec$  in the $V$ band.

{\em NGC 4736 = M 94.} NGC 4736 is the brightest galaxy of type Sa
in CVn\,I.  We resolve it into stars for the first time.
The WFPC2 was pointed at the galaxy periphery to avoid stellar
crowding. In the galaxy halo the CM diagram (Fig. 2) shows numerous of
RGB stars with $I$(TRGB) = 24.33$\pm 0.28$ mag, which yields distance of
4.66$\pm 0.59$ Mpc. Karachentseva \& Karachentsev (1998) carried out a
proper search for dwarf companions to NGC 4736 based on the POSS-II plates.
Surprisingly, they found no companions with a central surface brightness
brighter than 25 $^m/\sq\arcsec$ in the $B$ band within a radius of
$\sim3$ degrees or 230 kpc around this giant galaxy.
Such a pronounced degree of isolation of an Sa galaxy
situated in the middle of the CVn\,I cloud seems rather unusual.

{\em UGC 8308 = DDO 167.} This is an asymmetric irregular galaxy of
low surface brightness, becoming redder from its core to the periphery.
It has been resolved into stars by Tikhonov \& Karachentsev (1998), who
derived a distance of 3.7$\pm 0.7$ Mpc from the brightest star photometry.
The CMD (Fig. 2) shows blue and red stellar populations with the TRGB
position yielding a distance of 4.19$\pm 0.47$, which is in close agreement
with the previous estimate.

{\em UGC 8320 = DDO 168.} This irregular galaxy is located $14 \arcmin$
away from UGC 8308, forming a probable pair of dwarf galaxies. Bresolin
et al. (1993), Hopp \& Schulte-Ladbeck (1995), and Tikhonov \& Karachentsev
(1998) have resolved it into stars and estimated it distance to be
3.3 Mpc, 3.9 Mpc, and 4.0 Mpc, respectively. The WFPC2 photometry gives
a TRGB magnitude corresponding to a distance of 4.33$\pm 0.49$ Mpc.
The derived TRGB distances of UGC 8320 and UGC 8308 agree with each
other within the uncertainties.

{\em NGC 5204.} NGC 5204 is an irregular galaxy of Magellanic type,
which is located at the northern edge of the CVn\,I cloud. Its distance,
4.9$\pm 1.0$ Mpc, was estimated by Karachentsev et al.(1994) via the
brightest blue and red stars. The CM diagram (Fig. 2) shows a mixed
stellar population with a prominent RGB. The derived TRGB distance,
4.65$\pm 0.53$ Mpc, is in good agreement with the previous estimate.

{\em UGC 8833.} This blue irregular galaxy looks like a binary system
because of several regions with ongoing intense star formation.  
UGC 8833 is situated at the
eastern edge of the CVn\,I cloud. According to Makarova et al. (1998) its
distance derived from its brightest stars is 3.2$\pm 0.6$ Mpc. Our
photometry of the WFPC2 images yields a TRGB distance of 3.19$\pm 0.21$
Mpc, which confirms the previous distance estimate.

\section{ Properties of the Canes Venatici\,I cloud}

  Figure 3 presents the distribution of 223 galaxies within an area of 
$\alpha = 11^h 30^m$ to $13^h 40^m$, $\delta = 25\degr$ to $55\degr$ according to their radial
velocities with respect to the LG centroid. The data are taken from
the latest version of the Lyon Extragalactic Database (LEDA) prepared
by Paturel et al. (1996). The histogram shows a rather isolated peak
at $V_{LG} = 200 - 350$ km~s$^{-1}$, which is caused by galaxies in the CVn\,I cloud.
Another peak is seen in the range of 500 - 650 km~s$^{-1}$ and may correspond
to a more distant galaxy group aligned along the Supergalactic
equator (data on distances of these galaxies are yet unknown).
The distribution of 72 galaxies with  $V_{LG} < 550$ km~s$^{-1}$ is given in 
equatorial coordinates in Figure 4. The galaxies with $V_{LG} < 400$ km~s$^{-1}$
are shown by filled circles. Two brightest members of the CVn\,I cloud, NGC 4736
and NGC 4449, are indicated by filled squares.
Probable background galaxies with $400 < V_{LG} < 550$ km~s$^{-1}$ are
shown by crosses. The complete list of the galaxies is given in
Table 2. Its columns contain: (1) galaxy name, (2) equatorial coordinates
(1950.0), (3) apparent blue magnitude from the NED (uncorrected for
Galactic extinction),(4) morphological type,
(5) radial velocity with respect to the LG centroid, (6) distance
to the galaxy with indication of the used method: ``Cep'' -- via cepheids,
``RGB'' -- via the tip of red giant branch stars, ``SBF'' -- via surface
brightness fluctuations, ``BS'' -- from the luminosity of the brightest
stars, and ``GCLF'' -- via the globular cluster luminosity function. 
The last column presents the source of data on the distance. In
addition, we included in Table 2 an interacting galaxy pair NGC 5194/5195
with an accurate distance estimate and three nearby dwarf galaxies: DDO 187,
DDO 190, and KK 230, situated slightly to the east of the above indicated
boundary of the cloud.  The data given in Fig. 3, Fig. 4,
and Table 2 permit us to infer properties of the structure and kinematics
of the CVn\,I cloud.

  Judging from their distances and radial velocities, 34 galaxies may
be CVn\,I members. We distinguish their names in Table 2
with bold print. Among them there are 24 galaxies whose distances have
been measured with an accuracy of $\sim(10-15)$\%. For the other 10 probable
members of CVn\,I only rough distance estimates via the brightest stars
are known so far. Karachentsev \& Tikhonov (1994) claimed the typical error
of distance modulus for the ``BS''- method to be ~0.4 mag. However,
Rozanski \& Rowan-Robinson (1994) and some others considered the uncertainties
of this method
to be greater than 0.5 mag. In Table 2 there are 19 galaxies whose distances
have been measured at first via the brightest stars and then via TRGB.
Their distance moduli are given in Table 3.
Apart from three cases (UGC 7298, UGC 7577, and UGCA 290)
with $\Delta (m-M)$ larger than 1.0 mag, the mean difference of distance
moduli for the remaining 16 galaxies is $<(m-M)_{BS} - (m-M)_{RGB}> = -0.08 \pm 0.12$
mag, and the rms difference is 0.48 mag. As can be seen from
Table 2, apart from galaxies with rough distance estimates there are
also many galaxies whose distances have not yet been measured at all. For
some objects (PGC 38277, PGC 38685, PGC 91228 ) their unreliable
velocity estimates need to be checked. Due to the incompleteness of the
present set of observational data, our conclusions about the structure of the
CVn\,I cloud will have a preliminary character.

 Unlike the Local Group and the nearest groups around M81 and Cen A,
the galaxy complex in CVn\,I has no distinct dynamical center usually designated
by a giant early-type galaxy. We assume that the dynamical center of
CVn\,I lies between the two brightest cloud members, NGC 4736 and NGC 4449.
Their absolute magnitudes, $-19.69$ and $-18.37$ mag, are substantially fainter
than those of the Milky Way, M 31, M 81, and Centaurus A. As was
mentioned above, around the Sa galaxy NGC 4736 there is not any known
dwarf galaxy within 230 kpc. Such isolateness of NGC 4736 distinguishes
it from the brightest members of other groups. If the luminosity
of the brightest member of any group depends on the merging process of
surrounding dwarf galaxies, then the rate of the merging process in the CVn\,I
cloud was slow.

  The amorphous cloud CVn\,I differs essentially from more compact nearby
groups by its very sparse population of dSph galaxies.
Only one CVn\,I member, KK 166, may currently be considered to be a definitive
dSph galaxy.  
Another reddish LSB dwarf galaxy of regular shape, KDG 90, shows a strong
H\,{\sc i} flux, not typical of the dSphs. However,
KDG 90 is situated near the bright irregular galaxy NGC 4214, which may lead
to H\,{\sc i} flux confusion. Anyhow, the relative number of
dSphs in CVn\,I does not exceed 6\%, which also gives evidence of
low rate of interaction between the cloud galaxies if dSphs are primarily
the result of stripping in interactions.

  Comparing the luminosity function (LF) for field galaxies with the LF
for members of three nearest groups (LG + M 81 + Cen A),
Karachentsev et al. (2002c) noted an excess of very faint ($M_B > -12$ mag),
as well as giant ($M_B < -20$ mag) galaxies in the groups. The excess of
galaxies of extreme luminosities may be  understood if the primordial
LF grows on its bright and faint ends owing to ``cannibalism'' and ``debris''
left by galaxy interactions. In Fig. 5 we present the LF for 34 members
of the CVn\,I cloud together with the LFs for 38 field galaxies and 96
members of the three groups. As it follows from Fig. 5, the LF of the
CVn\,I seems to resemble the field LF more closely than the group LF. This
feature indicates once again that galaxy interactions do not necessarily 
exercise significant influence on the dynamical evolution of galaxies 
in the CVn\,I cloud.

  As was mentioned above, the boundary and the center position of
the CVn\,I cloud still remain uncertain. Based on the data of Fig. 4
and Table 2, one can speculate that the cloud is a superposition of several
groups populated by almost entirely irregular dwarf galaxies. In that sense, the
CVn\,I complex resembles another loose cluster of late-type galaxies in
Cancer (Bicay \& Giovanelli 1987) and nearby cloud of dIrr galaxies in
Ursa Major (Tully et al. 1996). Some groups in CVn\,I, for instance,
[NGC 4244, NGC 4395, UGC 7559, UGC 7605, IC 3687], [UGC 8215, UGC 8308,
UGC 8320, UGC 8331], and [UGC 8651, UGC 8760, UGC 8833] fit the definition
of groups of ``squelched'' galaxies introduced by Tully et al. (2002).
Luminous matter in such groups plays a negligible role in their
dynamical evolution.

\section{ Kinematics of the CVn\,I cloud}

 The sample of 34 possible members of the CVn\,I cloud, which are marked 
in Table 2 in bold print, are characterized by a mean distance
$ < D > = 4.1 \pm 0.2 $ Mpc and a mean radial velocity  $ < V_{LG} > = 286 \pm 9 $
km~s$^{-1}$. The ratio of these quantities yields 
$H(CVn\,I) = 70 \pm 4$ km~s$^{-1}$~Mpc$^{-1}$
as the local value of the Hubble constant,
which agrees well with its global value, $H_0 =
69 \pm 4$ (random) $\pm 6$ (systematic) km~s$^{-1}$~Mpc$^{-1}$ (Ferrarese et al. 2000).
In other words, within the uncertainties 
the CVn\,I cloud as a whole is at rest with respect to
the global cosmic flow within random errors.

  As a dynamical system, the CVn\,I cloud has the following integrated
parameters: a radial velocity dispersion $\sigma_v = 50$ km~s$^{-1}$,
a mean projected linear radius $< R_p >$ = 760 kpc, a mean harmonic
projected radius $< R_H >$ = 635 kpc, and an integrated luminosity of
$L_B = 2.15\cdot 10^{10} L_{\sun}$. Considering the CVn I cloud
to be in dynamical equilibrium and
applying the virial relation (Limber \& Mathews 1960)
$$M_{vir} = 3\pi N \cdot(N-1)^{-1} \cdot G^{-1} \cdot \sigma^2_v \cdot R_H,$$
where G is the gravitational constant and N is the number of group members, we
obtain the virial mass estimate  $$ M_{vir} = 3.6\cdot10^{12} M_{\sun} $$
and the virial mass-to-total luminosity ratio of 167 $M_{\sun}/L_{\sun}$.
Because half of the total luminosity of the CVn\,I cloud is emitted by
its brightest
galaxy, NGC 4736, we may consider formally the remaining cloud members
as companions to NGC 4736. Under these assumptions, the orbital mass
estimator is
 $$M_{orb} = (32/3\pi)\cdot G^{-1}\cdot (1- 2e^2/3)^{-1} <R_p\cdot \Delta V^2_r
>, $$
where $ e $ is the eccentricity of the Keplerian orbit, and $ R_i$ and $ \Delta V_i$
are the projected linear distance and radial velocity of a companion with respect
to NGC 4736. Adopting a mean eccentricity $e$=0.7, we derive
$M_{orb} = 1.9\cdot 10^{12} M_{\sun}$ and  $M_{orb}/L_{B} = 88$ $M_{\sun}/L_{\sun}$. Both estimates
are in satisfactory agreement with the mass-to-luminosity ratio,
$M_{vir}/L_{B} = 93$ $ M_{\sun}/L_{\sun}$ obtained by Tully (1987) for 22 members of
the CVn\,I cloud.

  Based on the present (incomplete) data on galaxy distances, we may
establish that the CVn\,I cloud extends in depth some 2.5 -- 3.5 Mpc,
namely, from $D_{min} = 2.5$ Mpc to $D_{max} = 5$ Mpc (via the TRGB method) or
even to 6 Mpc via the less reliable distance estimates from the brightest
stars. In the projection onto the sky the most distant CVn\,I members are
situated at $R_p ~\sim 1.4$ Mpc from the center. Hence, the CVn\,I cloud is a
system slightly elongated in space along the line of sight.

  It should be emphasized that such an extended complex of galaxies
with a low velocity dispersion (only 50 km~s$^{-1}$!) has not yet reached
the virialized state. The ``crossing time'' of the CVn\,I cloud defined as
$T_{cross} = < R_p>/ \sigma_v$ is ~15 Gyr, comparable to the cosmic
expansion time. Consequently, the derived estimates of the virial/orbital
mass should be used with great caution.

  What is the dynamical state of the CVn\,I cloud? Is it a semi-virialized
system or a structure taking part in the free Hubble flow? Figure 6
presents the distribution of galaxies in the CVn\,I region according to their
distances and radial velocities. Here the galaxies with accurate distance
estimates are shown by filled circles, and the galaxies with distances
known only via the brightest stars are indicated by crosses. Four luminous
galaxies with $M_B < -18$ mag are shown by filled squares. The straight
line that passes the foreground objects KK 230, DDO 187, and UGC 8508,
and the background galaxies UGCA 290, NGC 4258, and NGC 5194/95, fits a
Hubble constant H = 71 km~s$^{-1}$~Mpc$^{-1}$. The CVn\,I centroid position is
indicated by an open box whose sides correspond to the 1-$\sigma$ errors
of the mean distance and velocity. From the data we conclude that
the peculiar velocity of the centroid of the CVn\,I cloud does not exceed the
error of its determination, $\sim20$ km~s$^{-1}$. This result seems to be not
trivial, because of the existence of the Virgo-centric flow (Kraan-Korteweg,
1986) can generate significant deviations from the pure Hubble flow
on a scale of $\sim 5$ Mpc.

  The behavior of the members of the CVn\,I cloud in Fig. 6 reveals an interesting
feature: all the galaxies at the front of the cloud are situated
above the Hubble regression line. That may be caused by the differential
motion of the peripheral galaxies towards the cloud center at a velocity of
$\sim 65$ km~s$^{-1}$. In the case of spherical symmetry, a similar motion of
more distant cloud members towards its center (i.e., towards us) is expected
(Tonry et al. 2000). Unfortunately, the distances to galaxies at the back
of the cloud are known so far only with large errors, and the suspected
``back-flow'' effect turns out to be very noisy. But we believe that more
accurate distance measurements for a dozen galaxies on the back of
the cloud can easily clarify whether the backflow effect exists or not.

  Thus, returning to the question about the dynamical state of the CVn\,I cloud,
we suggest that the complex of predominantly irregular galaxies shows
some signs of deviation from the free Hubble expansion. But it seems to
be very far from the virialized state. Presumably evolving 
systems like the CVn\,I cloud, UMa cloud, and the Cancer cluster
are a common feature of the large scale structure of the universe.

  It should also be noted that in the CVn\,I region there are some galaxies
(UGC 7131, NGC 4150, KK 127, and UGC 7949) with radial velocities of
100 - 350 km~s$^{-1}$, but with distance estimates in the range of (10 - 20) Mpc.
These objects tend to be concentrated on the southern side of the cloud,
closer to the Virgo cluster. These galaxies may belong to the
Virgo cluster outskirts, and their low radial velocities may be caused by
large peculiar motions with respect to the Virgo core.

\section{Concluding remarks}
  As was noted above, the CVn\,I cloud is a scattered system
elongated towards the Local Group. Together with the LG and the loose
group in Sculptor, aligned along the line of sight, the CVn\,I cloud
consists of an amorphous filament of $\sim10$ Mpc in length. We estimated the
excess of the number of galaxies in the CVn\,I cloud to be $(N- \langle N \rangle)/ \langle N \rangle \sim 7$.
Because the complex is populated predominantly by dwarf galaxies,
the overdensity of the CVn\,I in its luminosity turns out to be lower,
$(\rho_{lum} - \langle \rho_{lum} \rangle)/ \langle \rho_{lum} \rangle \sim 4$. Here we adopted that the space
volume of the CVn\,I is 21 Mpc$^3$, and the mean luminosity of 1 Mpc$^3$
is $2\cdot 10^8 L_{\sun}$ (Bahcall et al. 1995).

  Remarkably, the mass estimates for the cloud, derived via orbital
and virial motions and being distributed over the 21 Mpc$^3$ volume,
yield an average density of dark matter $0.62 \rho_c$ and $1.17 \rho_c$,
respectively, where $\rho_c = 1.0\cdot 10^{-29}$ g/cm$^3$ is the critical density
with $H_o = 70$ km~s$^{-1}$~Mpc$^{-1}$.

  The large crossing time for the CVn\,I cloud (15 Gyr), the low content
of dSphs, the almost primordial shape of the LF indicate that the
CVn\,I complex is in a transient state not so far from the pure Hubble flow.
We hope that future accurate distance
measurements of about 20-30 members of the cloud in a fast snapshot survey
with the Advanced Camera will give a complete and reliable basis of
studying the dynamical evolution of the Canes Venatici complex.

\acknowledgements
Support for this work was provided by NASA through grant GO--08601.01--A
from the Space Telescope Science Institute, which is operated by the
Association of Universities for Research in Astronomy, Inc.,
under NASA contract NAS5--26555.
This work was partially supported by
 RFBR grant 01--02--16001 and DFG-RFBR grant 02--02--04012.
 D.G. gratefully acknowledges support from the Chile {\sl Centro de
Astrof\'\i sica} FONDAP No. 15010003. D.G. kindly acknowledges support
from ESO as a Visiting Astronomer which allowed him to work on this paper.

 The Digitized Sky Surveys were produced at the Space Telescope
Science Institute under U.S. Government grant NAG W--2166. The
images of these surveys are based on photographic data obtained
using the Oschin Schmidt Telescope on the Palomar Mountain and the UK
Schmidt Telescope. The plates were processed into the present
compressed digital form with permission of these institutions.

 This project made use of the NASA/IPAC Extragalactic Database (NED),
which is operated by the Jet Propulsion Laboratory, Caltech, under
contract with the National Aeronautics and Space Administration.

{}
\onecolumn
\newpage
\begin{table}
\scriptsize{
\caption{New distances to galaxies in the Canes Venatici cloud}
\begin{tabular}{lccccrcccrrr} \\ \hline
 Name &   RA (1950) Dec &    a x b & $ B_t$ & $A_b$ & $(V-I)_T $ &  T & $V_{LG}$ & I(TRGB)& $ (m-M)_o $ &  D & $(V-I)_{-3.5}$  \\
      & $^{hh mm ss}\;\;\;\;\;   \degr\degr \;\arcmin\arcmin \;\arcsec\arcsec$ & $ \arcmin$ &   mag&  $ A_i$ & $\pm \sigma$ &  & km~s$^{-1}$ &  mag  &  mag &  Mpc & [Fe/H] \\
\hline
U6541 & 113045.2 493043 & 1.4x 0.8& 14.23& 0.08 & 0.53     & 10 & 304  &23.94  &27.95 & 3.89 & 1.39 \\
      &                 &         &      & 0.04 & $\pm$0.07 &    &      & $\pm$0.28  & $\pm$0.29 & $\pm$0.47 & -1.54 \\
      &                 &         &      &      &          &    &      &       &      &      \\
N3738 & 113304.4 544758 & 2.6x 1.9& 12.13& 0.05 & 0.62    & 10 & 305  &24.42  &28.45 & 4.90 & 1.48 \\
      &                 &         &      & 0.02 & $\pm$0.10 &    &      & $\pm$0.24  & $\pm$0.25 & $\pm$0.54 & -1.27 \\
      &                 &         &      &      &          &    &      &       &      &      \\
N3741 & 113325.2 453343 & 2.0x 1.1& 14.3 & 0.10 & 0.67    & 10 & 264  &23.41  &27.41 & 3.03 & 1.37 \\
      &                 &         &      & 0.05 & $\pm$0.10 &    &      & $\pm$0.21  & $\pm$0.23 & $\pm$0.33 & -1.61 \\
      &                 &         &      &      &          &    &      &       &      &      \\
KK109 &  114433.5 435659&  0.6x 0.4& 18.62& 0.08& 0.80     &  10&  241 & 24.26 & 28.27&  4.51 & 1.15\\
      &                 &         &      &  0.04& $\pm$0.10 &    &      &  $\pm$0.15 &  $\pm$0.17&  $\pm$0.34 & -2.54\\
      &                 &         &      &      &          &    &      &       &      &      \\
N4150 &  120801.2 304054&  2.3x 1.6& 12.45& 0.08& 1.20     &  -1&  198 &   -   &   -  & 20: & - \\
      &                 &         &      &  0.04& $\pm$0.01 &    &      &       &      &   &   \\
      &                 &         &      &      &          &    &      &       &      &      \\
U7298 &  121400.6 523018&  1.1x 0.6& 15.95& 0.10& 0.45     &  10&  255 & 24.12 & 28.12&  4.21 & 1.32\\
      &                 &         &      &  0.05& $\pm$0.26 &    &      &  $\pm$0.15 &  $\pm$0.17&  $\pm$0.32 & -1.80\\
      &                 &         &      &      &          &    &      &       &      &      \\
N4244 &  121459.8 380506& 19.4x 2.1& 10.67& 0.09& 0.89     &   6&  255 & 24.25 & 28.26&  4.49 & 1.32\\
      &                 &         &      &  0.04& $\pm$0.07 &    &      &  $\pm$0.22 &  $\pm$0.24&  $\pm$0.47 & -1.80\\
      &                 &         &      &      &          &    &      &       &      &      \\
N4395 &  122320.8 334922& 13.2x11.0& 10.61& 0.07& 0.73     &   9&  315 & 24.30 & 28.32&  4.61 & 1.45\\
      &                 &         &      &  0.03& $\pm$0.11 &    &      &  $\pm$0.28 &  $\pm$0.29&  $\pm$0.57 & -1.36\\
      &                 &         &      &      &          &    &      &       &      &      \\
U7559 &  122437.1 372509&  3.2x 2.0& 14.12& 0.06& 0.48     &  10&  231 & 24.42 & 28.44&  4.87 & 1.29\\
DDO126&                 &         &      &  0.03& $\pm$0.10 &    &      &  $\pm$0.24 &  $\pm$0.26&  $\pm$0.55 & -1.91\\
      &                 &         &      &      &          &    &      &       &      &      \\
N4449 &  122545.1 442215&  6.2x 4.4&  9.83& 0.08& 0.63     & 10&  249 & 24.11 & 28.12 &  4.21 & 1.43 \\
      &                 &         &      &  0.04& $\pm$0.02 &   &      &  $\pm$0.26 &  $\pm$0.27 &  $\pm$0.50 & -1.42 \\
      &                 &         &      &      &          &   &      &       &      &      \\
U7605 &  122611.0 355940&  1.1x 0.8& 14.76& 0.06& 0.61     & 10&  317 & 24.21 & 28.23&  4.43 & 1.21\\
      &                 &         &      &  0.03& $\pm$0.10 &   &      &  $\pm$0.27 &  $\pm$0.28&  $\pm$0.53 & -2.26\\
      &                 &         &      &      &          &   &      &       &      &      \\
IC3687&  123950.8 384633&  3.4x 3.0& 13.75& 0.09& 0.57     & 10&  385 & 24.29 & 28.30&  4.57 & 1.32\\
DDO141&                 &         &      &  0.04& $\pm$0.26 &   &      &  $\pm$0.22 &  $\pm$0.24&  $\pm$0.48 & -1.80\\
      &                 &         &      &      &          &   &      &       &      &      \\
KK166 &  124649.5 355305&  1.7x 1.0& 17.62& 0.06& 1.20     & -3&   -  & 24.36 & 28.38&  4.74 & 1.27 \\
      &                 &         &      &  0.03& $\pm$0.2  &   &      &  $\pm$0.34 &  $\pm$0.34&  $\pm$0.69 & -2.00\\
      &                 &         &      &      &          &   &      &       &      &      \\
N4736 &  124832.3 412328& 11.2x 9.1&  8.74& 0.08& 1.19     &  2&  353 & 24.33 & 28.34&  4.66 & 1.69\\
M94   &                 &         &      &  0.04& $\pm$0.01 &   &      &  $\pm$0.28 &  $\pm$0.29&  $\pm$0.59 & -0.84\\
      &                 &         &      &      &          &   &      &       &      &      \\
U8308 &  131110.0 463511&  1.1x 0.6& 15.53& 0.04& 0.69     & 10&  243 & 24.08 & 28.11&  4.19 & 1.37\\
DDO167&                 &         &      &  0.02& $\pm$0.26 &   &      &  $\pm$0.25 &  $\pm$0.26&  $\pm$0.47 & -1.61\\
      &                 &         &      &      &          &   &      &       &      &      \\
U8320 &  131216.6 461101&  3.6x 1.4& 12.73& 0.07& 0.52     & 10&  273 & 24.16 & 28.18&  4.33 & 1.28\\
DDO168&                 &         &      &  0.03& $\pm$0.26 &   &      &  $\pm$0.25 &  $\pm$0.26&  $\pm$0.49 & -1.96\\
      &                 &         &      &      &          &   &      &       &      &      \\
N5204 &  132743.8 584032&  5.0x 3.0& 11.73& 0.05& 0.75     &  9&  341 & 24.31 & 28.34&  4.65 & 1.32\\
      &                 &         &      &  0.02& $\pm$0.01 &   &      &  $\pm$0.26 &  $\pm$0.27&  $\pm$0.53 & -1.80\\
      &                 &         &      &      &          &   &      &       &      &      \\
U8833 &  135236.0 360500&  0.9x 0.8& 15.15& 0.05& 0.68     & 10&  285 & 23.49 & 27.52&  3.19 & 1.26\\
      &                 &         &      &  0.02& $\pm$0.10 &   &      & $\pm$0.12 &  $\pm$0.15&  $\pm$0.21 & -2.04\\
      &                 &         &      &      &          &   &      &       &      &      \\
\hline
\end{tabular}}
\end{table}

\begin{table}
\caption{Galaxies in the Canes Venatici region with $ V_{LG} < 550 $ km~s$^{-1}$}
\begin{tabular}{lrrrrll} \\ \hline
Name    &   RA (B1950) Dec & $ B_t$ & T & $ V_{LG}$ & Distance &    Reference   \\
 \hline
{\bf U6541} & 113045.9 493052 & 14.23 & 10 & 304 & 3.89 RGB &  present paper        \\
{\bf N3738} & 113304.4 544758 & 12.13 & 10 & 305 & 4.90 RGB &  present paper        \\
{\bf N3741} & 113325.2 453343 & 14.3  & 10 & 264 & 3.03 RGB &  present paper        \\
{\bf KK109} & 114433.5 435659 & 18.62 & 10 & 241 & 4.51 RGB &  present paper        \\
{\bf U6817} & 114816.8 390931 & 13.44 & 10 & 248 & 2.64 RGB &  Karachentsev \&,2002c \\
{\bf N4068} & 120129.7 525201 & 13.19 & 10 & 290 & 5.2  BS  &  Makarova \&,1997      \\
     N4080 & 120218.6 271616 & 14.28 & 10 & 519 &          &                       \\
     Mrk757& 120242.9 310801 & 14.80 & 0:& 551 &          &                       \\
     P38286& 120250.2 283839 & 15.36 & 10 & 527 &          &                       \\
     U7131 & 120639.4 311106 & 15.50 & 8 & 226 &14.   BS  &  Makarova \&,1998      \\
     P38685& 120724.5 364248 & 15.5  & 9:& 341 &          &                       \\
     N4144 & 120728.2 464407 & 12.16 & 6 & 319 & 9.7  BS  &  Karachentsev \&,1998  \\
     N4150 & 120801.4 304047 & 12.45 & -2 & 198 &20: GCLF  &  present paper        \\
{\bf N4163} & 120937.5 362651 & 13.93 & 10 & 164 & 3.6  BS  &  Tikhonov \&,1998      \\
     KK127 & 121051.0 301159 & 15.61 & 10 & 105 &          &                       \\
{\bf N4190} & 121113.5 365440 & 13.52 & 10 & 234 & 3.5  BS  &  Tikhonov \&,1998      \\
{\bf KDG90} & 121227.1 362948 & 15.40 & -1 & 283 & 2.86 RGB &  Karachentsev \&,2002c \\
{\bf N4214} & 121308.2 363619 & 10.24 & 10 & 295 & 2.94 RGB &  Maiz-Apellaniz \&,2002\\
     P39228& 121318.5 523955 & 15.3  & 10 & 245 &          &                       \\
{\bf U7298} & 121400.6 523018 & 15.95 & 10 & 255 & 4.21 RGB &  present paper        \\
{\bf N4244} & 121459.8 380506 & 10.67 & 6 & 255 & 4.49 RGB &  present paper        \\
     N4248 & 121523.0 474109 & 13.12 & 9:& 544 &          &                       \\
     N4258 & 121629.4 473453 &  9.10 & 4 & 507 & 7.28 SBF &  Tonry \&,2001         \\
     U7356 & 121641.0 472202 & 15.10 & 10 & 330:&          &  (HI flux confusion?)\\
     U7369 & 121708.1 300938 & 14.70 & -1 & 198 &          &                       \\
     U7408 & 121847.5 460520 & 13.35 & 10 & 515 &          &                       \\
     IC3247& 122043.8 291015 & 15.25 & 8 & 539 &          &                       \\
     IC3308& 122247.7 265929 & 15.17 & 7 & 277 &          &                       \\
     KK144 & 122258.0 284533 & 16.5  & 10 & 453 &          &                       \\
{\bf N4395} & 122320.8 334922 & 10.61 & 9 & 315 & 4.61 RGB &  present paper        \\
{\bf UA281} & 122350.5 484607 & 15.15 & 10 & 349 & 5.7  BS  &  Makarova \&,1997      \\
{\bf U7559} & 122437.1 372509 & 14.12 & 10 & 231 & 4.87 RGB &  present paper        \\
{\bf U7577} & 122515.4 434613 & 12.84 & 10 & 240 & 2.54 RGB &  Karachentsev \&,2002c \\
{\bf N4449} & 122545.1 442215 &  9.83 & 10 & 249 & 4.21 RGB &  present paper        \\
     U7599 & 122600.8 373035 & 14.98 & 9 & 291 & 6.9  BS  &  Makarova \&,1998      \\
{\bf U7605} & 122611.0 355940 & 14.76 & 10 & 317 & 4.43 RGB &  present paper        \\
\hline
\end{tabular}
\end{table}
\begin{table}
\begin{tabular}{lrrrrll} \\ \hline
     N4460 & 122620.0 450827 & 12.26 & -1 & 542 & 9.59 SBF &  Tonry \&,2001         \\
     KK149 & 122625.8 422715 & 15.01 & 10 & 446 &          &                       \\
     U7639 & 122728.4 474822 & 14.13 & 10 & 446 & 8.0  BS  &  Makarova \&,1998      \\
     KK151 & 122758.0 431039 & 15.8  & 9 & 479 &          &                       \\
     N4485 & 122805.1 415833 & 12.32 & 10 & 530 &          &                       \\
     U7699 & 123021.5 375352 & 13.17 & 8 & 514 &          &                       \\
{\bf U7698} & 123024.9 314853 & 13.15 & 10 & 321 & 6.1  BS  &  Makarova \&,1998      \\
     UA290 & 123456.0 390112 & 15.74 & 10 & 484 & 6.70 RGB &  Crone \&,2002         \\
{\bf UA292} & 123613.3 330229 & 16.1  & 10 & 306 & 3.1  BS  &  Makarova \&,1998      \\
{\bf IC3687}& 123950.8 384633 & 13.75 & 10 & 385 & 4.57 RGB &  present paper        \\
     KK160 & 124135.0 435615 & 17.   & 10 & 346 &          &                       \\
    FGC1497& 124435.2 325521 & 16.   & 9 & 521 &          &                       \\
     U7949 & 124435.9 364457 & 15.12 & 10 & 351 &10.   BS  &  Makarova \&,1998      \\
{\bf KK166} & 124649.5 355305 & 17.62 & -3 &     & 4.74 RGB &  present paper        \\
     U7990 & 124801.0 283726 & 16.2  & 10 & 495 &          &                       \\
{\bf N4736} & 124832.3 412328 &  8.74 & 2 & 353 & 4.66 RGB &  present paper        \\
{\bf U8024} & 125139.3 272510 & 14.17 & 10 & 355 & 4.3  BS  &  Makarova \&,1998      \\
{\bf IC4182}& 130329.9 375223 & 12.41 & 9 & 356 & 4.70 Cep &  Sandage \&,1982       \\
{\bf U8215} & 130550.4 470524 & 16.07 & 10 & 297 & 5.6  BS  &  Makarova \&,1997      \\
     N5023 & 130957.9 441813 & 12.82 & 6 & 476 & 5.4  BS  &  Sharina \&,1999       \\
{\bf U8308} & 131110.8 463504 & 15.53 & 10 & 243 & 4.19 RGB &  present paper        \\
     KK191 & 131124.0 421831 & 18.2  & 10 & 429 &          &                       \\
{\bf U8320} & 131216.6 461101 & 12.73 & 10 & 273 & 4.33 RGB &  present paper        \\
     U8331 & 131320.3 474537 & 14.61 & 10 & 345 & 8.2  BS  &  Karachentsev \&,1998  \\
{\bf N5204} & 132743.8 584032 & 11.73 & 9 & 341 & 4.65 RGB &  present paper        \\
     N5194 & 132749.7 472932 &  8.57 & 5 & 555 & 7.7  mem &  pair with N5195      \\
     N5195 & 132752.4 473132 & 10.45 & -1 & 558 & 7.66 SBF &  Tonry \&,2001         \\
{\bf U8508} & 132847.1 551002 & 14.40 & 10 & 186 & 2.56 RGB &  Karachentsev \&,2002c \\
     N5229 & 133158.5 481016 & 14.51 & 7 & 460 & 5.1  BS  &  Sharina \&,1999       \\
{\bf N5238} & 133242.6 515209 & 13.8  & 8 & 345 & 5.2  BS  &  Karachentsev \&,1994  \\
{\bf U8638} & 133658.5 250144 & 14.47 & 10 & 273 & 2.3: BS  &  Makarova \&,1998      \\
{\bf U8651} & 133744.2 405931 & 14.7  & 10 & 272 & 3.01 RGB &  Karachentsev \&,2002c \\
{\bf U8833} & 135236.0 360500 & 15.15 & 10 & 285 & 3.19 RGB &  present paper        \\
     KK230 & 140501.5 351809 & 17.9  & 10 & 125 & 1.90 RGB &  Grebel \&,2001        \\
     DDO187& 141338.6 231713 & 14.38 & 10 & 174 & 2.50 RGB &  Aparicio \&,2000      \\
     DDO190& 142248.7 444506 & 13.25 & 10 & 266 & 2.79 RGB &  Karachentsev \&,2002c \\
\hline
\end{tabular}
\end{table}

\begin{table}
\caption{Distance moduli for 19 galaxies in Canes Venatici derived
from their brightest stars (BS) and RGB stars.}
\begin{tabular}{lccc} \\ \hline
Galaxy  & $ (m-M)_{BS}$ & $ (m-M)_{RGB}$ & difference \\
\hline
UGC 6541 &   27.73 &   27.95 &    -0.22  \\
NGC 3738 &   27.73 &   28.45 &    -0.72   \\
NGC 3741 &   27.70 &   27.41 &     0.29   \\
UGC 6817 &   27.97 &   27.11 &     0.86   \\
NGC 4214 &   28.06 &   27.34 &     0.72   \\
UGC 7298 &   29.67 &   28.12 &    (1.55)  \\
NGC 4244 &   28.28 &   28.26 &     0.02   \\
NGC 4395 &   28.13 &   28.32 &    -0.20   \\
UGC 7559 &   27.97 &   28.44 &    -0.47   \\
UGC 7559 &   28.53 &   28.44 &     0.09   \\
UGC 7577 &   28.42 &   27.02 &    (1.40)   \\
NGC 4449 &   27.33 &   28.12 &    -0.79   \\
UGC 7605 &   28.24 &   28.23 &     0.01   \\
UGCA 290 &   27.20 &   29.13 &   (-1.93)  \\
IC  3687 &   27.37 &   28.30 &    -0.93   \\
UGC 8308 &   27.85 &   28.11 &    -0.26   \\
UGC 8320 &   28.01 &   28.18 &    -0.17   \\
NGC 5204 &   28.46 &   28.34 &     0.12   \\
UGC 8651 &   27.66 &   27.39 &     0.27   \\
UGC 8833 &   27.53 &   27.52 &     0.01   \\
\hline
\end{tabular}
\end{table}

\begin{figure*}
\centering
\vspace{5mm}

\includegraphics[width=12cm]{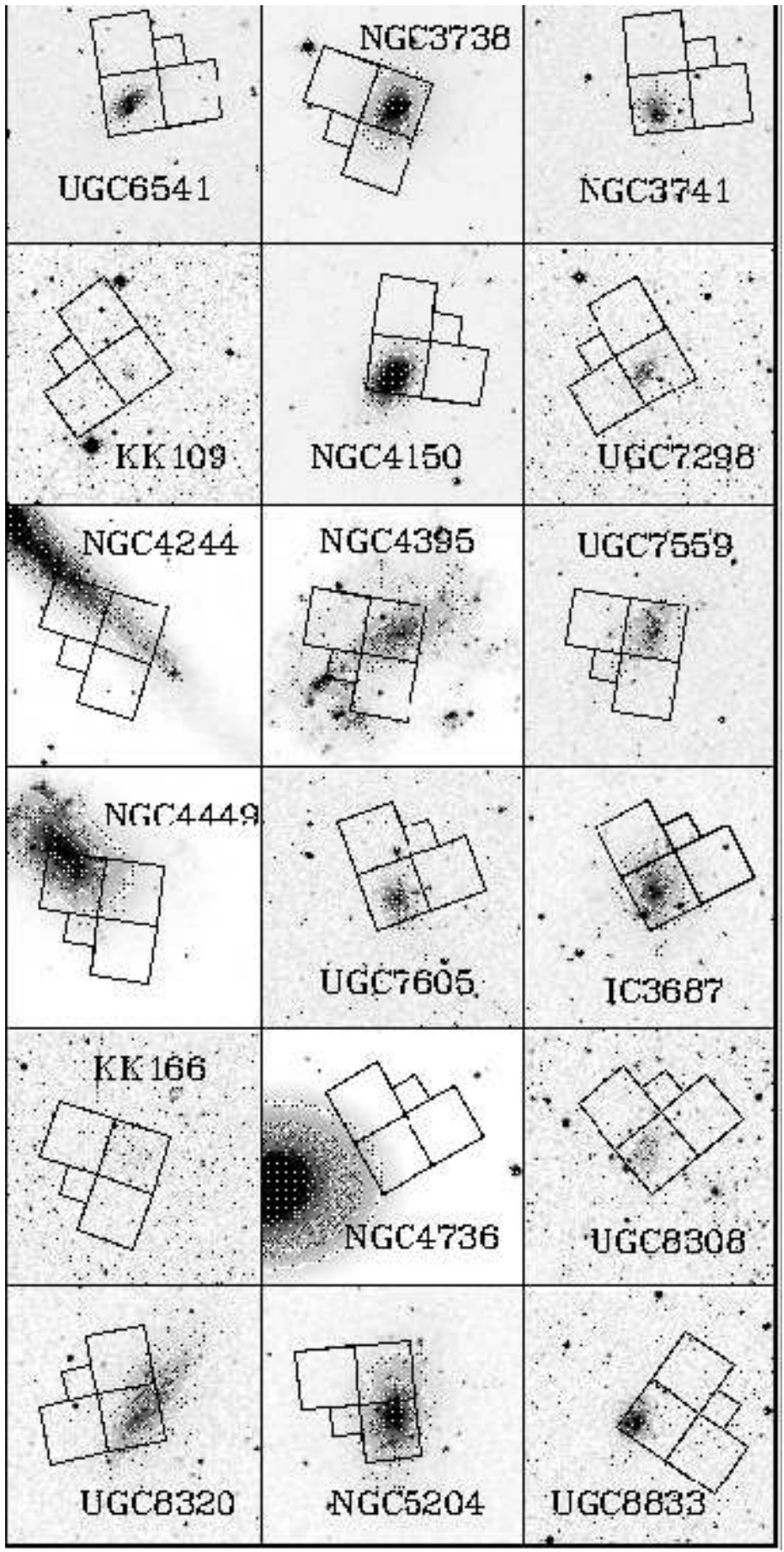}
\vspace{5mm}

\caption{ Digital Sky Survey images of 18 galaxies in the Canes Venatici I
cloud. The field size is 6$\arcmin$, North is up and East is left.
The HST WFPC2 footprints are superimposed.}
\end{figure*}
\clearpage

\begin{figure*}
\centering
\vspace{-10mm}

\vspace{-10mm}
\caption{{\bf Top}: WFPC2 images of 18 galaxies: UGC 6541, NGC 3738, NGC 3741,
KK 109, NGC 4150, UGC 7298, NGC 4244, NGC 4395, UGC 7559, NGC 4449, UGC 7605,
IC 3687, KK 166, NGC 4736, UGC 8308, UGC 8320, NGC 5204, and UGC 8833,
produced by combining the two 600s exposures obtained through the F606W
and F814W filters. The arrows point to the North and the East.
{\bf Bottom left}: The color-magnitude diagrams from the WFPC2 data for
the 18 galaxies in the CVn\,I cloud.
{\bf Bottom right}: the Gaussian-smoothed $I$-band luminosity function restricted
to red stars (top), and the
output of an edge-detection filter applied to the luminosity function
for the 18 galaxies studied here.}
\end{figure*}

\begin{figure*}
\centering
\vspace{5mm}

\includegraphics[width=12cm,angle=-90]{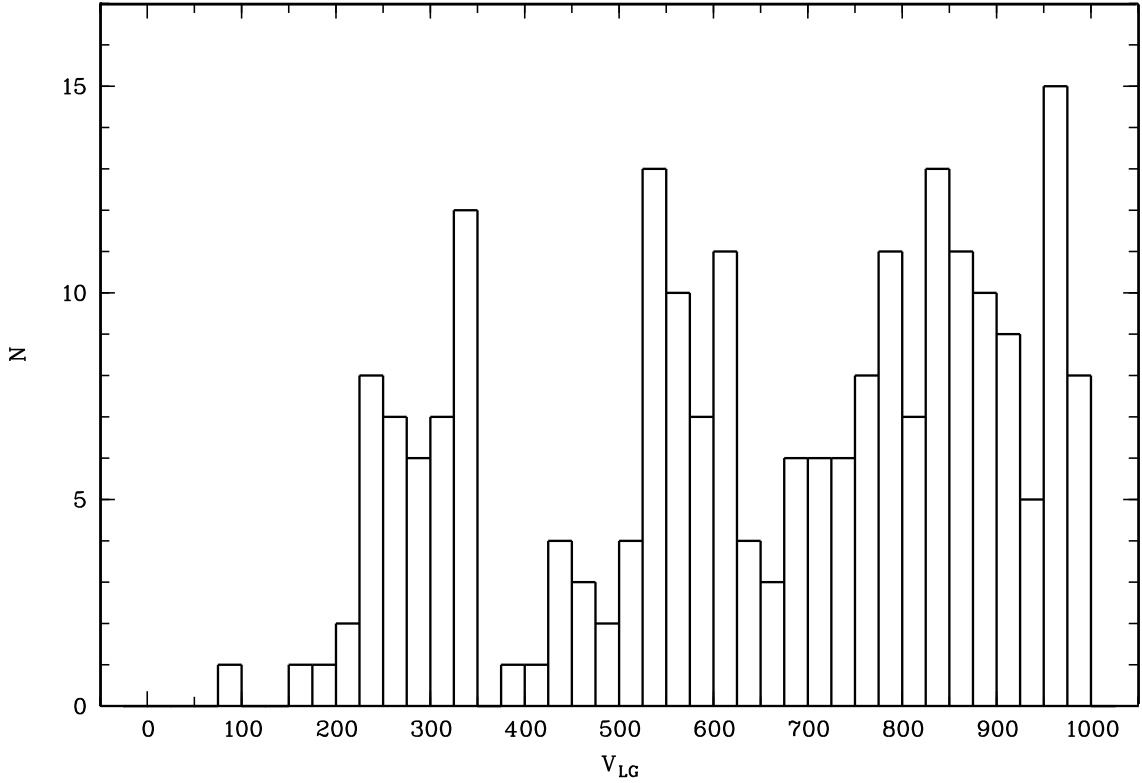}
\vspace{5mm}
\caption{ Radial velocity distribution of 223 galaxies in the CVn\,I region.}
\end{figure*}

\begin{figure*}
\centering
\vspace{5mm}

\includegraphics[width=13cm,angle=-90]{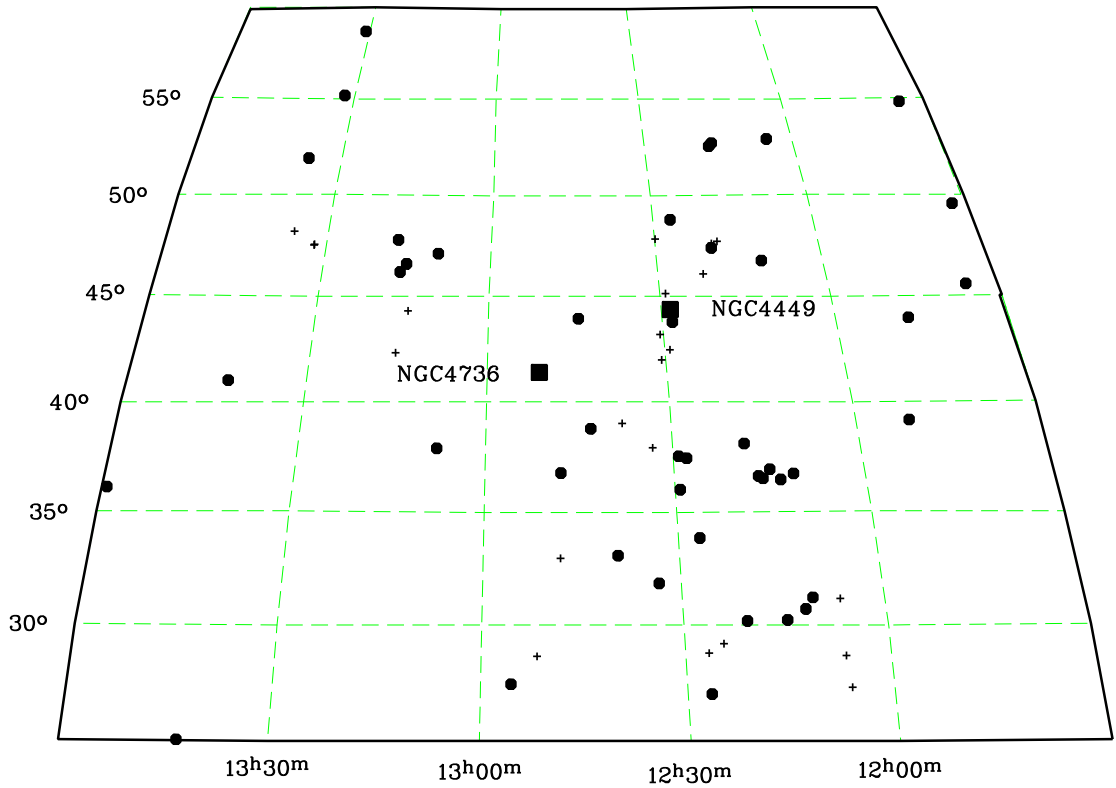}
\vspace{5mm}
\caption{ The distribution of 72 galaxies with corrected radial velocities
$V_{LG} < 550$ km~s$^{-1}$ in the Canes Venatici constellation in equatorial
coordinates. The galaxies with $V_{LG} < 400$ km~s$^{-1}$ and
$> 400$ km~s$^{-1}$ are indicated by filled circles and crosses,
respectively. The two brightest galaxies, NGC 4736 and NGC 4449,
are shown by filled  squares.}
\end{figure*}

\begin{figure*}
\centering
\vspace{5mm}

\includegraphics[width=12cm,angle=-90]{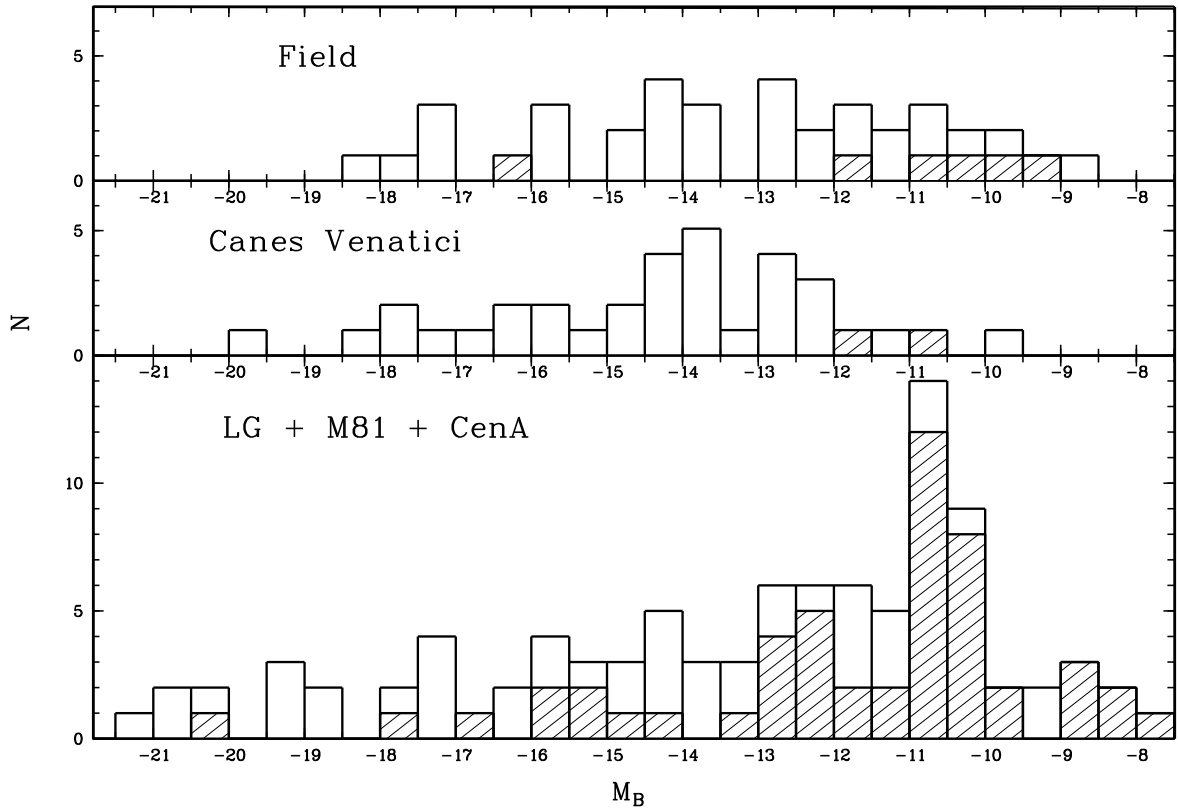}
\vspace{5mm}
\caption{ The luminosity function of 34 CVn\,I cloud galaxies (middle histogram),
38 nearby field galaxies (upper histogram) and 96 galaxies situated in the LG,
the M81 group, and the CenA group. E and dSph galaxies are shaded.}
\end{figure*}

\begin{figure*}
\centering
\vspace{5mm}

\includegraphics[width=12cm,angle=-90]{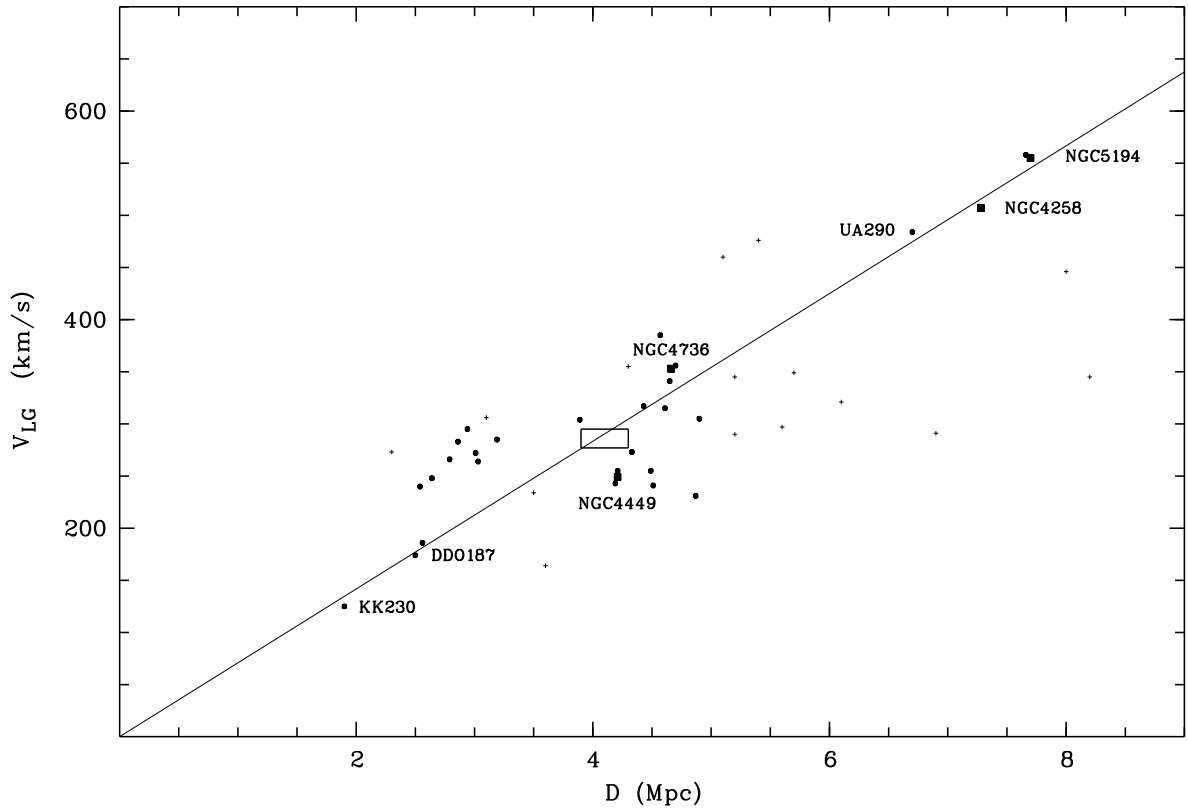}
\vspace{5mm}
\caption{ Velocity-distance relation for galaxies in the CVn\,I region. The
galaxies with accurate distance estimates are indicated by filled circles,
The most luminous of them are shown as filled squares. The galaxies with
distances estimated via the brightest stars are marked by crosses. The
straight line corresponds to the Hubble constant $H_o = 71$ km~s$^{-1}$~Mpc$^{-1}$.
Position of the CVn\,I centroid is shown as an open box with sides equal
to 1-$\sigma$ errors of the mean distance and velocity.}

\end{figure*}

\end{document}